\documentclass[runningheads]{llncs}
\usepackage[T1]{fontenc}
\usepackage{graphicx}
\usepackage[nolist]{acronym}
\usepackage{xspace}
\usepackage[capitalise]{cleveref}
\usepackage{siunitx}
\usepackage{url}
\usepackage{booktabs,dcolumn}
\usepackage{multirow}

\makeatletter
\newcolumntype{B}[3]{>{\boldmath\DC@{#1}{#2}{#3}}c<{\DC@end}}
\makeatother

\def\etal{\emph{et al.}\xspace}

\def\ie{i.\,e.\xspace}

\begin{document}
\title{Domain and Content Adaptive Convolutions for Cross-Domain Adenocarcinoma Segmentation}
%
%
\author{Frauke Wilm \inst{1,2,*}\and
Mathias Öttl\inst{1,2} \and
Marc Aubreville\inst{3} \and
Katharina Breininger\inst{1,4}}
\authorrunning{F. Wilm et al.}
%
\institute{Department Artificial Intelligence in Biomedical Engineering, Friedrich-Alexander-Universität (FAU) Erlangen-Nürnberg, Erlangen, Germany \and
Pattern Recognition Lab, FAU Erlangen-Nürnberg, Erlangen, Germany \and
Flensburg University of Applied Sciences, Flensburg, Germany \and
Center for AI and Data Science, Julius-Maximilians-Universität Würzburg, Würzburg, Germany}

\maketitle              
\begin{abstract}
Recent advances in computer-aided diagnosis for histopathology have been largely driven by the use of deep learning models for automated image analysis. While these networks can perform on par with medical experts, their performance can be impeded by out-of-distribution data. The \acf{cosas} challenge aimed to address the task of cross-domain adenocarcinoma segmentation in the presence of morphological and scanner-induced domain shifts. In this paper, we present a U-Net-based segmentation framework designed to tackle this challenge. Our approach achieved segmentation scores of \num{0.8020} for the cross-organ track and \num{0.8527}  for the cross-scanner track on the final challenge test sets, ranking it the best-performing submission.  

\end{abstract}
\let\thefootnote\relax\footnotetext{*~corresponding author: \email{frauke.wilm@fau.de}}
\section{Introduction}
With the development of designated slide scanners, traditional pathology has experienced a shift towards digital pathology, allowing histologic samples to be diagnosed on a computer screen rather than under an optical microscope. This transition has not only facilitated global expert collaboration but also enabled the application of machine learning models for computer-aided diagnosis. These models have demonstrated human-like performance across various diagnostically relevant tasks~\cite{srinidhi2021deep}, such as mitotic figure assessment or tumor segmentation. However, their performance can be substantially impacted by out-of-distribution samples~\cite{wilm2023mind}. In histopathology, such samples may originate from the differing morphologies of various organs or from the visual discrepancies introduced by scanners from different vendors~\cite{aubreville2023comprehensive}. The \acf{cosas} challenge, held as a satellite event of the International Conference on \ac{miccai} 2024, addressed the task of robust cross-domain adenocarcinoma segmentation in histology samples. This manuscript presents a methodological approach to solving this challenge, which ranked among the top-performing submissions on the preliminary test set and achieved the highest segmentation scores on the final test set.

\section{Challenge Tasks and Datasets}
The \ac{cosas} challenge comprises two cross-domain histology datasets for two challenge tracks. Each dataset is composed of 290 \acp{roi} of human adenocarcinoma samples routinely stained with \ac{he}. Each \ac{roi} has an average size of 1500~x~1500 pixels and tumor lesions were manually labeled with contour annotations. For the challenge, the organizers divided each dataset into train, preliminary test, and final test sets with 180-20-90 images respectively.

\subsubsection{Cross-Organ Dataset (Task 1)} The images of the cross-organ dataset were all digitized with the  TEKSQRAY SQS-600P scanner and comprised images from six different organs. Of these six domains, only three were included in the training dataset (gastric adenocarcinoma, colorectal adenocarcinoma, and pancreatic ductal adenocarcinoma) and the others remained undisclosed to the challenge participants. The preliminary test set was composed of images from four different organs (including two training domains) and the final test set of images from all six organs. In each dataset, a uniform sample distribution across included domains was ensured.

\subsubsection{Cross-Scanner Dataset (Task 2)} The \acp{roi} of the cross-scanner dataset were all obtained from invasive breast carcinoma samples of no special type acquired with six distinct scanning systems. Of these six scanners, only three were included in the training dataset (TEKSQRAY SQS-600P, KFBIO KF-PRO-400, and  3DHISTECH PANNORAMIC 1000) and the others remained undisclosed to the challenge participants. The preliminary test set was composed of images from four scanners and the final test set of images from all six scanning systems. In each dataset, a uniform sample distribution across included domains was ensured. \\

\noindent The challenge participants were not allowed to use any additional datasets for model development and pre-trained models were restricted to conventional, non-medical image datasets. However, the participants were allowed to combine the datasets of both challenge tracks and submit the same model to both tracks. 

\section{Methods}
\begin{figure}[!ht]
\includegraphics[width=\textwidth]{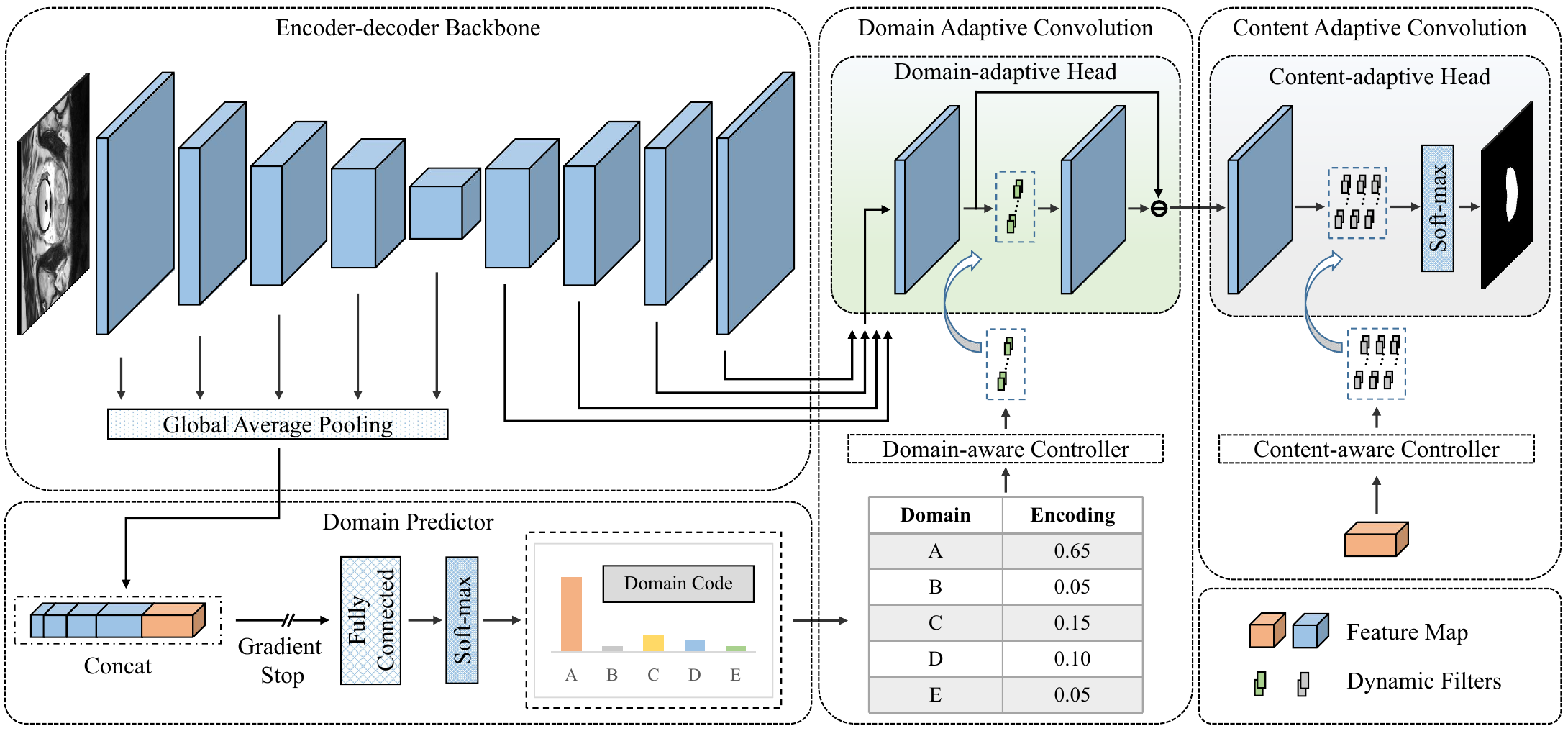}
\caption{Overview of the \acf{dcac} module. Figure adapted from Hu \etal~\cite{hu2022domain}.} \label{fig:architecture}
\end{figure}

For the task of cross-domain adenocarcinoma segmentation, we adopted nnU-Net~\cite{isensee2021nnu}, which utilizes a dataset fingerprint to automatically determine the optimal U-Net~\cite{ronneberger2015u} configuration for a given task. Furthermore, we experimented with the \ac{dcac}-based multi-source domain generalization approach, proposed by Hu \etal~\cite{hu2022domain}, visualized in \cref{fig:architecture}. \Ac{dcac} utilizes two modules, \ie a \ac{dac} module and a \ac{cac} module, and can be incorporated into any standard encoder-decoder segmentation network. 

\noindent For the \ac{dac} module, the multiscale encoder feature maps are global average pooled, concatenated, and passed on to a domain predictor, which computes an individual domain encoding, \ie a vector with probabilistic values for belonging to a source domain. Images from seen source domains are ideally one-hot encoded, whereas domain encodings for unseen target domain images are interpolated from the source domains. This domain encoding then controls the parameters of dynamic filtering kernels in the domain-adaptive head, resulting in domain-adaptive convolutions. Besides domain-adaptive convolutions, \ac{dcac} also employs content-adaptive convolutions to adapt the architecture to variations across individual samples. The \ac{cac} module uses the pooled feature map of the last encoder layer to control the parameters of dynamic filtering kernels in the content-adaptive head, which is applied sequentially to the output of the domain-adaptive head.

\subsubsection{Implementation Details} 
We utilized the default experiment planner of nnU-Net to find the best model configurations. The model was trained with \num{250} minibatches per epoch and a batch size of \num{2}. We employed the default minibatch sampling method of nnU-Net. For optimization, we used the SGD optimizer with Nesterov momentum ($\mu$~=~\num{0.99}) and a polynomial learning rate scheduler (starting at 0.01). For segmentation training, we combined Dice and cross-entropy loss functions. For the \ac{dcac} module, we followed the implementations by Hu \etal~\cite{hu2022domain}, and a standard cross-entropy loss was used for optimizing the domain classifier. The default training period of nnU-Net is \num{1000} epochs, after which we observed convergence of the segmentation loss. For the \ac{dcac} module, however, the domain classifier accuracy still improved, so we trained the model for \num{2500} epochs in total. During training, five-fold cross-validation was used and model selection was guided by the best performance on the validation set, assessed using the exponential moving average of the segmentation Dice score (and additionally domain classification accuracy for the \ac{dcac} module). During inference, we applied model ensembling across all folds and test-time augmentation, consistent with the standard nnU-Net implementation. All code for model training and evaluation is publicly available in our GitHub repository: \url{https://github.com/DeepMicroscopy/nnUNet}.  

\section{Evaluation and Results}
We first tested both approaches (the baseline nnU-Net and the \ac{dcac} module) for out-of-distribution performance by training them on the training set of task 1 and evaluating them on the training set of task 2. In these cross-domain experiments, the domain encoder of the \ac{dcac} module was trained with three domains and demonstrated superior performance to the nnU-Net baseline (fourth row in \cref{tab:results}). We then randomly selected ten images of each domain as a hold-out validation set and combined the remaining images of both datasets into a single training set. For this experiment, the domain encoder of the \ac{dcac} module was trained with six domains. The results of these in-domain experiments are reported in rows 1 and 4 of \cref{tab:results}. Here, both methods performed comparably. 

For challenge submission, a Docker-based system was used, restricting participants from direct access to the test set. Participants were allowed up to five submissions per task on the preliminary test set and only one submission per task on the final test set. The final challenge ranking was based on a weighted average of the segmentation performance on both test sets, with weights of 0.2 for the preliminary and 0.8 for the final test set. Segmentation performance was evaluated as the average of the Dice and Jaccard scores. We used all of the 360 training images for training the final models and submitted them for evaluation on the preliminary test set. Again, both models performed comparably. We hypothesize that the combined dataset with six domains may have required better optimization of the domain classifier’s hyperparameters to demonstrate superior performance. We plan to explore this further after the final challenge deadline.

As only one submission was allowed on the final test set, we decided to submit the slightly better-performing nnU-Net baseline, which ranked first place in both challenge tracks with segmentation scores of \num{0.8020} and \num{0.8527}, respectively.

\begin{table}
\caption{Segmentation performance assessed as the average of the Dice and Jaccard scores. Evaluation for the internal validation sets was conducted by the authors, whereas the preliminary and final test set performance was evaluated by the challenge organizers.}\label{tab:results}
\setlength\tabcolsep{0pt}
\begin{tabular*}{\linewidth}{@{\extracolsep{\fill}}clll }
\toprule
& & nnU-Net & nnU-Net + DCAC\\
\midrule
\multirow{3}{*}{\rotatebox[origin=c]{90}{task 1}} &
in-domain (internal) & 0.8249 & \textbf{0.8412} \\
& preliminary test set & \textbf{0.7776} & 0.7690\\
& final test set & 0.8020 & n.a. \\
\midrule
\multirow{4}{*}{\rotatebox[origin=c]{90}{task 2}} &
cross-domain (internal) & 0.7574 & \textbf{0.7896} \\
& in-domain (internal) & \textbf{0.8413} & 0.8338 \\
& preliminary test set & \textbf{0.8858} & 0.8829 \\
& final test set & 0.8527 & n.a. \\
\bottomrule
\end{tabular*}
\end{table}

\begin{credits}
\subsubsection{\ackname} The authors gratefully acknowledge the scientific support and HPC resources provided by the Erlangen National High Performance Computing Center (NHR@FAU) of the Friedrich-Alexander-Universität Erlangen-Nürnberg (FAU) under the NHR project b209cb. NHR funding is provided by federal and Bavarian state authorities. NHR@FAU hardware is partially funded by the German Research Foundation (DFG) – 440719683. F.W. M.Ö., and K.B. acknowledge support by the German Research Foundation (DFG) project 460333672 CRC1540 EBM. K.B. further acknowledges support by d.hip campus - Bavarian aim in form of a faculty endowment.

\subsubsection{\discintname}
The authors have no competing interests to declare that are
relevant to the content of this article.
\end{credits}

%
%
%
\bibliographystyle{splncs04}
\bibliography{mybibliography}

\begin{acronym}
\acro{cosas}[COSAS]{Cross-Organ and Cross-Scanner Adenocarcinoma Segmentation}
\acro{dac}[DAC]{domain-adaptive convolution}
\acro{cac}[CAC]{content-adaptive convolution}
\acro{dcac}[DCAC]{Domain and Content Adaptive Convolution}
\acro{he}[HE]{Hematoxylin~\&~Eosin}
\acro{miccai}[MICCAI]{Medical Image Computing and Computer Assisted Intervention}
\acro{roi}[ROI]{region of interest}
\acrodefplural{roi}{regions of interest}
\end{acronym}

\end{document}